\documentclass[10pt]{iopart}
\usepackage[]{graphicx}
\begin{document}


\title{Vibrational signature of broken chemical order in a GeS$_2$ glass: a
molecular dynamics simulation} 

\author{S\'ebastien Blaineau and Philippe Jund}
\address{
Laboratoire de Physicochimie de la Mati\`ere Condens\'ee , 
Universit\'e Montpellier 2, \\Place E. Bataillon, Case 03, 
34095 Montpellier, France
}

\begin{abstract}
Using density functional molecular dynamics simulations, 
we analyze the broken chemical order in a GeS$_2$ glass and its
impact on the dynamical properties of the glass through the in-depth  
study of the vibrational eigenvectors. We find homopolar bonds and
the frequencies of the corresponding modes are in agreement with 
experimental data. Localized S-S modes and 3-fold coordinated sulfur 
atoms are found to be at the origin of specific Raman peaks whose origin
was not previously clear. Through the ring size statistics we find, 
during the glass formation, a conversion of 3-membered rings into larger 
units but also into 2-membered rings whose vibrational signature is 
in agreement with experiments.
\end{abstract}
 
\section{INTRODUCTION}
Dynamical properties of germanium disulfide glasses have been 
extensively studied in the last two decades using Raman
 and InfraRed (IR) spectroscopy 
\cite{nemanich,sugai,bose,boolchand,perakis,bychkov}. However, the 
analysis of these experimental results can be extremely complicated and 
the determination of the origin of the vibrational modes observed in Raman 
and IR spectra can therefore be hypothetical \cite{boolchand,perakis,bychkov}. 
Theoretical calculations based on molecular dynamics simulations can be a 
very interesting tool in order to eliminate these uncertainties since 
they allow the atomic displacements at the origin of each vibrational
 mode to be investigated. However, the importance of charge transfers in 
chalcogenide glasses \cite{transfer} such as GeS$_2$ requires the use of an 
{\em ab-initio} model which can be extremely costly in time if the size of
the system is not relatively small. In a previous study we have investigated  
a glassy GeS$_2$ sample containing 96 particles \cite{blaineau}, 
using a non self-consistent {\em ab-initio} code based on the 
Sankey-Niklewski scheme called FIREBALL \cite{fireball}. 
The good agreement of the results with
the available experimental data showed the excellent quality of this model
for the study of chalcogenide systems. However, the relative small size of 
the sample did not allow a great variety of structural defects 
inherent to amorphous systems and made a statistical determination 
of the broken chemical order in g-GeS$_2$ difficult. 
Experimental spectroscopic studies 
of these glasses \cite{boolchand} have shown in addition that the structural 
disorder manifests itself clearly in the vibrational properties of 
germanium disulfide glasses and it is therefore interesting to directly 
investigate the origin of the corresponding vibrational modes.\\
Here we propose a study of a larger sample (258 particles) of glassy GeS$_2$ 
using the same model, in order to study the statistics of the broken chemical 
order of these materials, and to analyze in detail the 
vibrational properties of amorphous GeS$_2$ at ambient temperature. 
In section II we will briefly present the model which has been used (more 
details can be found in the original paper \cite{fireball}). Subsequently, 
we will describe the structural disorder of our amorphous sample at 
short- and medium-range in section III. In section IV we will present a 
detailed study of the vibrational properties of the system in order to 
understand which atomic displacements are responsible of the modes present 
in the Vibrational Density of States (VDOS). Finally in section V we will 
summarize the major conclusions 
of our work.\\

\section{MODEL}
The code we have used is a first-principles type molecular-dynamics program 
called FIREBALL96, which is based on the local orbital electronic structure 
method developed by Sankey and Niklewski \cite{sankey}. The foundations of 
this model are Density Functional Theory (DFT) \cite{dft} within the 
Local Density Approximation (LDA) \cite{lda}, and the non-local 
pseudopotential scheme \cite{pseudo}. The use of the non-self-consistent 
Harris functional \cite{harris}, with a set of four atomic orbitals 
(1$s$ and 3$p$) per site that vanish outside a cut-off radius of $5a_0$ 
(2.645~\AA ) considerably reduces the CPU time.\\
As said earlier, this model has given excellent results for many different 
chalcogenide systems over the last ten years \cite{blaineau,drabold,junli}.
In the present work, we have melted a $\alpha$-GeS$_2$ crystal containing 
258 atoms in a cubic cell of 19.21~\AA~ at 2000K for 60 ps 
(24000 timesteps) in order 
to obtain an equilibrium liquid system (standard periodic boundary conditions have been used). 
We have then quenched our sample at a quenching 
rate of 6.8*10$^{14}$ K/s through the glass transition (the simulated glass 
transition temperature is close to 1200 K). Finally we have relaxed this system at ambient 
temperature for 600~ps. 
The results shown in this article have been averaged over this period.

\section{STRUCTURAL DISORDER}
The broken chemical order of amorphous systems can be revealed through a detailed study of 
short- and medium-range order in the structure of the sample. 
The study of the radial pair correlation function g(r) defined as\\
\begin{eqnarray} 
g(r)_{\alpha-\beta}=\frac{V}{4\pi r^2N_{\alpha}dr}dn_{\beta}
\end{eqnarray}
for a given $\alpha-\beta$ pair, is a standard way in order to reveal
 the existence of bond defects in glassy samples. We report in Fig.1 the 
results obtained for pairs of the same nature: Ge-Ge and S-S. The peaks 
corresponding to second neighbour distances, at 
3.64~\AA~for S-S pairs, and at 2.91~\AA~and 3.41~\AA~for Ge-Ge pairs 
(depending of whether the intertetrahedral connections are edge or corner-sharing), 
have already been analyzed in our previous work on the small sample 
\cite{blaineau} (in addition to the g(r) for the Ge-S pairs that does not
change significantly in the large sample). 
However in the 258-particle sample, we observe in both graphs a small peak 
at short distances (2.42~\AA~ for Ge-Ge pairs and 2.23~\AA~for S-S pairs), 
corresponding to nearest-neighbour bonds. These peaks indicate the presence 
of homopolar bonds, albeit in small quantities, in amorphous GeS$_2$ at 
ambient temperature. Several studies have been done on this topic, and 
recently two experimental studies have yielded conflicting results 
concerning the presence or not of homopolar bonds in stoichiometric 
glassy GeS$_2$ \cite{cai,petri}. Contrarily to Cai et al. who used 
Raman scattering \cite{cai}, Petri et $al.$ found no 
evidence for such bonds in g-GeS$_2$ in a neutron diffraction study 
\cite{petri}. It should be noted that the small concentration of these bonds, 
as we see them in our simulation, could explain why they can not be detected 
in the experimental structure factor since their effect is not strong enough to be 
clearly distinguished.\\ 
The Ge-Ge homopolar bonds appear in ethane-like units, where
each germanium is linked to another germanium and to three sulfur atoms. 
An illustration of this kind of structural unit is 
shown in Fig.2(a). The S-S bonds as we see them in our simulation can 
appear in two ways: either one of the sulfur atom is non-bridging (linked to one 
germanium Fig.2(b)), or both are non-bridging (Fig.2(c)). In both cases 
the S-S bond length is 2.26~\AA. We have found no S-S bonds containing 
two bridging sulfur atoms.\\ 
The structural disorder can also appear between atoms of different nature,
when the number of first neighbours of a given particle is different from 
what we can expect in a crystal. In stoichiometric $\alpha$-GeS$_2$, one 
Ge atom is linked to four S atoms, and each S atom is linked to two Ge atoms. 
The ``ideal'' corresponding amorphous system is then a glassy network of 
GeS$_4$ tetrahedra, linked together by one or two S atoms depending on the 
type of connection between the tetrahedra. A tetrahedron which is involved 
in an edge-sharing link is called an edge-sharing tetrahedron, and their 
presence can be experimentally detected using high-resolution neutron 
diffraction measures. In our simulation we find that 46$\%$ of the tetrahedra
 are edge-sharing in glassy GeS$_2$, which is extremely close to the experimental value 
of 44$\%$ determined by Bychkov et al.\cite{bychkov}.\\
The broken chemical order can also be seen at short length scales through the 
analysis of the bond defects of the amorphous sample. If we study the sulfur atoms of our 
system, we find that only 68$\%$ of them are ''correctly'' linked to two Ge atoms 
(the Ge-S bond length is 2.23~\AA\ (experiment: 2.21~\AA \cite{expe})).  
We find that 14.53$\%$ of them are connected to only one Ge atom, 
and are therefore called terminal sulfur (or non-bridging). The interatomic distance 
becomes then smaller than usual, and reaches the value of 2.12 \AA. This decrease in 
the bond length, whether the sulfur atoms are bridging or non-bridging has been 
studied experimentally and theoretically and similar results have been
 found \cite{ribes,popovic,cluster}.\\
The reduced coordination of the sulfur atoms is balanced by the presence of 
13.95$\%$ of 3-fold coordinated sulfur, which are connected 
to three germanium atoms instead of two (Fig.2(d)). 
In these configurations, the bond length increases to 2.29~\AA. All these values are reported in Table I.\\

\begin{tabular}{cccc}
\multicolumn{4}{c}{TABLE I. First neighbours of the S atoms in a g-GeS$_2$ system containing 258 particles}\\
\hline
\hline
\multicolumn{4}{c}{S}\\
No. of first Ge neighbours ~~~~ &~~~~Amount~~~~~~  &Percentage&Bond length\\
\hline
1&25&14.53$\%$&2.12 \AA\\
2&117&68$\%$&2.23 \AA\\
3&24&13.95$\%$&2.29 \AA\\
\hline
Homopolar bonds S-S&6&3.52$\%$&2.26 \AA\\
\hline\hline
Coordination number&n$_{tot}$=2.005&n$_{Ge}$=1.97&n$_{S}$=0.035\\
\hline\hline
\end{tabular}\\
\vspace{0.5cm}\\
If we now focus on the germanium atoms, we see that 95.36$\%$ of 
them are found in usual GeS$_4$ tetrahedra, with four nearest S  neighbours. 
Only 1.16$\%$ of the Ge atoms are involved in Ge(S$_{1/2})_3$ units and 1.16$\%$ in 
Ge(S$_{1/2})_2$ units. We can therefore say that the bond defects in g-GeS$_2$ 
are more specific to the sulfur atoms, since the coordination of germanium is 
extremely close to that of crystalline $\alpha$-GeS$_2$.\\
It can be seen that the Ge-S bond length increases in Ge(S$_{1/2})_3$ and Ge(S$_{1/2})_2$ 
units, and reaches the value of 2.43 \AA~ (Table II).
\vspace{0.5cm}\\
\begin{tabular}{cccc}
\multicolumn{4}{c}{TABLE II. First neighbours of the Ge atoms 
in a g-GeS$_2$ system containing 258 particles}\\
\hline
\hline
\multicolumn{4}{c}{Ge}\\
No. of first neighbours S~~~~~ &~~~~Amount~~~~~~  &Percentage&Bond length\\
\hline
2&1&1.16$\%$&2.43 \AA\\
3&1&1.16$\%$&2.43 \AA\\
4&82&95.36$\%$&2.23 \AA\\
\hline
Homopolar bonds Ge-Ge&2&2.32$\%$&2.42 \AA\\
\hline\hline
Coordination number&n$_{tot}$=3.953&n$_{S}$=3.93&n$_{Ge}$=0.023\\
\hline\hline
\end{tabular}\\
\vspace{0.5cm}\\
	An alternative way to analyze the chemical order at medium range can 
be obtained through the ring size  statistics. We define the size $n$ of a ring as the 
number of germanium atoms present in the shortest {\em closed} path of alternating 
Ge-S bonds. Therefore a $n$-membered ring consists of $2n$ alternating Ge-S bonds. 
In crystalline $\alpha$-GeS$_2$, the ring size can only be equal to 2 or 3. 
Their respective proportion is 1:2, which means that 33$\%$ 
of them are 2-membered rings and 66$\%$ of them are 3-membered rings. However, in amorphous 
g-GeS$_2$, these rings may have more variable sizes, and their statistics is a signature of 
medium range disorder. We report in Table III the respective amount and percentage 
of rings for $n$=2,....,8 in our glassy sample (no $n$-membered ring 
with $n>8$ was found in our 258-particles system).\\
We note that the percentage of 2-membered rings (38$\%$) is extremely close to that of 
crystalline $\alpha$-GeS$_2$. However, many 3-membered rings 
have disappeared in the amorphous structure, to become $n$-membered rings, with $n=2$ 
and $3<n\leq 8$. These units were probably broken during the
process that led to the glassy structure, allowing for the formation of mainly 
larger-sized rings. To our knowledge, neither experimental results nor other 
simulation data have been published on the ring size statistics in 
g-GeS$_2$.\\ 

\begin{tabular}{ccc}
\multicolumn{3}{c}{~~~~~~~~TABLE III.  Ring statistics in a g-GeS$_2$ system containing 258 particles}\\
\hline\hline
~~~~~~~~~~~~~~~~~~Ring size~~~~~~~~~~~~~~&No. of rings&Percentage\\
\hline
2&22&38$\%$\\
3&27&46.5$\%$\\
4&1&1.7$\%$\\                              
5&2&3.5$\%$\\
6&2&3.5$\%$\\
7&3&5.1$\%$\\
8&1&1.7$\%$\\
\hline\hline
\end{tabular}\\
\vspace{0.5 cm}\\
Once we can describe in detail the chemical ``defects'' in our glassy GeS$_2$ sample
(with respect to the crystal), we can move on to the study of the vibrational
properties of this system in order to analyze the impact of these defects 
on the dynamics, impact which is supposed to be important according to
experimental studies \cite{boolchand}. 

\section{DYNAMICAL PROPERTIES}

The study of the dynamical properties of g-GeS$_2$ can be done through the calculation of the 
Vibrational Density of States (VDOS), which can be measured experimentally by inelastic 
neutron diffraction spectroscopy. Even though 
Raman and IR measurements have been performed in chalcogenide glasses, no neutron diffraction 
experiments have ever been performed on g-GeS$_2$ glasses to our knowledge.\\
We calculate the VDOS of glassy g-GeS$_2$ through the diagonalization of D, the dynamical 
matrix of the system given by:
\begin{eqnarray}
D(\phi_i,\phi_j)=\frac{\partial^2 E(\phi_i,\phi_j)}{\partial\phi_i\partial\phi_j} , ~~\phi=x,y,z
\end{eqnarray}
for two particles $i$ and $j$. Fig.3 shows the calculated VDOS for two g-GeS$_2$ samples 
containing respectively 96 and 258 particles. In the larger sample (the topic of the present 
study) we notice a widening of the optical band leading to an excess of modes (compared to the 
small sample) between 250 and 300 cm$^{-1}$ 
and above 450 cm$^{-1}$. We will see later that these modes arise from structural defects 
that are not present in the small sample (as seen previously \cite{blaineau}) but in order 
to analyze the VDOS and to determine the nature of the atomic displacements responsible of 
the vibrational modes, we first calculate the Phase Quotient \cite{bell1}, 
which is defined as:
\begin{eqnarray}
PQ(\omega)=\frac{\Sigma_m \vec{e}^{\:1}_{m}(\omega).\vec{e}_m^{\:2}(\omega)}{\Sigma_m |\vec{e}_m^{\:1}(\omega).\vec{e}_m^{\:2}(\omega)|}
\end{eqnarray}
where the summation is done over all first-neighbour bonds in the system.
$\vec{e}_m^{\:1}(\omega)$ and $\vec{e}_m^{\:2}(\omega)$ are the eigenvectors for eigenvalue 
$\omega$ of the 2 particles involved in the $m^{th}$ first-neighbour bond. 
If the motion of particles {\em 1} and {\em 2} 
is parallel for all first-neighbour bonds, then PQ($\omega$) is equal to +1. If, on the 
contrary, the motion is antiparallel for all pairs, then PQ($\omega$) has a value of -1. 
Fig.4  represents the calculated Phase Quotient for our glassy GeS$_2$ sample. It can be 
seen that the first band in the VDOS at low frequencies is 
caused by acoustic-like modes, in which a particle vibrates almost in phase with its first 
neighbours. These modes involve therefore extended interblock vibrations. The second 
band in the VDOS arises from optic-like modes, in which the eigenvectors of first 
neighbours are mostly opposed to each other, 
leading mainly to intrablock vibrations. This is consistent with the usual assignment 
of the acoustic and optic character of the two main bands in the VDOS.\\
Another interesting tool in the analysis of the vibrations of a system is the Stretching character $S_c$ \cite{zotov}, which is defined as: 
\begin{eqnarray}
S_c=\frac{\Sigma_m (\vec{e}_m^{\:1}(\omega)-\vec{e}_m^{\:2}(\omega)).\vec{d}_m}{\Sigma_m |\vec{e}_m^{\:1}(\omega)-\vec{e}_m^{\:2}(\omega)|}
\end{eqnarray}
The summation is performed over all first neighbour links, and $\vec{d}_m$ is the unit vector between the two particles of the $m^{th}$ bond.
The stretching character of g-GeS$_2$ is shown in Fig.5.
We can see that at low frequencies, the vibrational modes arise from bending-like displacements, since S$_c$ is close to 0. On the other hand, high frequency modes in the optic band can be attributed to stretching-like vibrations, in which the eigenvectors are almost parallel to the direction along the first-neighbour bonds, and S$_c$ is close to 1.\\
In order to measure the localization of the modes, we calculate the participation ratio P$_r$ \cite{bell2}:
\begin{eqnarray}
P_r=\frac{(\Sigma_{i=1}^N|\vec{e}_i(\omega)|^2 )^2}{N\Sigma_{i=1}^N|\vec{e}_i(\omega)|^4}
\end{eqnarray}
where the summation is performed over the N particles of the sample. If the mode corresponding to 
the eigenvalue $\omega$ is delocalized and all atoms vibrate with equal amplitudes, then P$_r$($\omega$) will be close to 1. On the contrary, if the mode is strongly localized, then 
P$_r$($\omega$) will be close to 0. The results are shown in Fig.6(a) and we can see that the 
localized modes are mainly present in the zone between the acoustic and the optic bands and also at 
higher frequencies at the upper limit of the optic band. In this work, we will focus on these localized modes, not present in the small sample.\\
In order to determine which particles are involved in a given mode, we define the center of gravity $\vec{r}_g(\omega)$ of each mode of eigenvalue $\omega$, and the corresponding ``localization'' length L \cite{zotov}, as
\begin{eqnarray}
\vec{r}_g(\omega)=\frac{\Sigma_{i=1}^N\vec{r}_i|\vec{e}_i(\omega)|^2/m_i}{\Sigma_{i=1}^N |\vec{e}_i(\omega)|^2/m_i}
\end{eqnarray}
and
\begin{eqnarray}
L(\omega)=\sqrt{\frac{\Sigma_{i=1}^N|\vec{r}_i-\vec{r}_g(\omega)|^2|\vec{e}_i(\omega)|^2/m_i}{\Sigma_{i=1}^N|\vec{e}_i(\omega)|^2/m_i}}
\end{eqnarray}
where $\vec{r}_i$ and $m_i$ are respectively the position and the atomic mass of particle $i$. Periodic boundary conditions are of course taken into account in these calculations. The localization length 
(Fig.6(b)) represents the spatial localization of a given mode. It is a length beyond which the 
amplitude of the atomic vibrations decreases significantly. Its maximal value is half that of the 
box size which is in this case equal to 9.6~\AA~. For each eigenvalue, a sphere of radius L centered at $\vec{r}_g(\omega)$ will define the zone in which 
the vibration is located. All the particles present inside this sphere will be considered as 
involved in the 
vibration. This scheme will allow us to determine the atomic displacements responsible of each localized mode.\\
 A very controversial aspect in the study of the dynamics of g-GeS$_2$ is the interpretation of the modes 
located in the range close to 250 cm$^{-1}$, since the origin of the peak observed in Raman spectroscopy has 
never been clearly understood. Experiments on Ge-enriched  Ge$_x$S$_{1-x}$ systems, with different 
concentrations of germanium, have been performed and the feature at 250 cm$^{-1}$ was found to increase with the 
Ge concentration\cite{boolchand,bychkov,lucovsky2}. This led to the hypothesis that this feature is due to tetrahedral units containing less than 4 sulfur atoms, such as Ge(S$_{1/2})_3$ or Ge(S$_{1/2})_2$ units 
\cite{lucovsky2}. In comparable studies these vibrations have been assigned to Ge-Ge homopolar 
bonds \cite{lucovsky1,jackson}, or to distorted rocksalt units \cite{boolchand}. This feature, 
at 250 cm$^{-1}$, was also found to increase in intensity under high pressure conditions \cite{weinstein}. In our simulation 
we find that these vibrational modes are localized around 3-fold coordinated sulfur atoms, 
with a corresponding localization length approximately equal to the first-neighbour distance between the 
central S atom and the 3 Ge first neighbours. This mode is therefore extremely localized, and the 3 Ge atoms 
are almost frozen in comparison to the central S that vibrates between them. This finding is consistent  
with all the experimental results mentioned above. Since Ge-rich systems contain more of 
these bond defects it can explain why the feature at 250 cm$^{-1}$ in Raman spectra increases in Ge-rich 
compositions. Under high pressure conditions, where Ge-rich and S-rich zones have been proposed to appear \cite{lucovsky2}, the same explanation holds.
\\
We find that two different zones, at 200 cm$^{-1}$ and 440 cm$^{-1}$, are localized around edge-sharing units 
(2-membered rings). The high-frequency feature, at 440 cm$^{-1}$, has been analyzed in several  works 
\cite{boolchand,bychkov,jackson}, and its origin is indeed due to the vibrations of edge-sharing units. 
These modes are caused by symmetric breathing-like displacements as depicted in Fig.7(a). The other feature, 
at 200 cm$^{-1}$, which is not yet clearly understood, was found to increase in Ge-rich compositions \cite{bychkov}. 
We find that these modes are centered around edge-sharing units as well and are due to the vibrations 
shown in Fig.7(b). Since the concentration of edge-sharing tetrahedra increases significantly with 
Ge concentration \cite{bychkov}, this could explain why this feature is more prominent in Ge-rich 
compositions.\\ 
The homopolar bonds (Ge-Ge and S-S) also have a contribution in the Vibrational Density of States. 
We find that the vibrations of ethane-like units, presenting Ge-Ge bonds, have a contribution at 270 cm$^{-1}$ 
which is close to the value determined experimentally (260 cm$^{-1}$) \cite{bychkov} and 
theoretically (250 cm$^{-1}$) \cite{jackson}.\\
We find the vibrational modes containing S-S homopolar bonds at 486 cm$^{-1}$, which is exactly the value 
found in Raman experiments on sulfur-rich samples \cite{boolchand}. These modes have been attributed to 
the S-S bonds present in S$_8$ rings or S$_n$ chains, which appear in S-rich 
compositions \cite{sugai,lucovsky2,jackson,s8exp}. However, in our stoichiometric GeS$_2$ glass we find 
no presence of these chains and the corresponding localization length of the mode (see Fig.6(b)) is extremely 
small ($\approx$1.5 \AA). We can therefore argue that the feature at 486 cm$^{-1}$ in Raman spectra is not 
necessarily caused by {\em extended} S$_8$ rings or S$_n$ chains, but exists as soon as localized S-S homopolar bonds 
are present. These homopolar bonds are only present in the large sample which explains why we observe in the VDOS
an excess of modes around $\approx$ 270 cm$^{-1}$ and 486 cm$^{-1}$ compared to the VDOS of the small
system (Fig.3).
\\ 
Another region has been extensively studied over the past 20 years, concerning the so-called A$_1$ 
and A$_{1_c}$ modes, at 340 cm$^{-1}$ and 370 cm$^{-1}$ \cite{sugai,boolchand,bychkov,jackson}. 
Their presence in Raman spectra has been determined as arising from the symmetric stretch of 
Ge(S$_{1/2})_4$ 
tetrahedra, in which the central germanium atom is frozen. The ''companion'' mode A$_{1_c}$ at 370 cm$^{-1}$ has 
been proposed to be due to similar vibrations in edge-sharing tetrahedra. The A$_1$ and A$_{1_c}$ modes are
mainly Raman-active, but cannot be identified clearly in the VDOS. Nevertheless since we have 
previously shown that in our simulation the other modes have frequencies very close to the 
experimental values, we have
decided to analyze the modes between  340 and 370 cm$^{-1}$. We indeed find tetrahedral 
symmetric-stretching vibrations between 340 cm$^{-1}$ and 360 cm$^{-1}$, as represented in Fig.8(a)  
We also find at 375 cm$^{-1}$, symmetric-stretching modes concerning edge-sharing tetrahedra, 
which are usually 
assigned to the A$_{1_c}$ modes (Fig.8(b)). In addition, many symmetric stretching vibrations of 
tetrahedra 
containing bond defects are present between the A$_1$ and A$_{1_c}$ zones. This can explain why the feature 
at 340 cm$^{-1}$ in Raman spectra does not disappear in Ge-rich compositions \cite{lucovsky2,boolchand}, 
even though the usual Ge(S$_{1/2})_4$ tetrahedral units are no longer present in such systems. 
\section{CONCLUSION}
Using DFT based MD simulations, we have analyzed the structural disorder of a 
glassy GeS$_2$ sample containing 258 particles and its impact on the vibrational spectrum.
Compared to a previous study on a smaller system \cite{blaineau}, we find here the presence of homopolar bonds 
(even though in extremely small concentration) which {\em a priori} permits the disagreement between 
two recent experimental studies to be resolved \cite{cai,petri}. The signature of these bonds in the VDOS at 270~cm$^{-1}$ and 
 486~cm$^{-1}$ is in agreement with previous experimental and theoretical results. Nevertheless, the feature at 
486 cm$^{-1}$, previously assigned to S-S vibrations in S$_8$ rings or S$_n$ chains that can be found in S-rich 
compositions, is not necessarily due to such medium sized units, but can arise from by S-S vibrations at 
short distances. We also propose an alternative explanation of the feature at 250 cm$^{-1}$ in the Raman spectrum, 
showing that the corresponding  modes are centered around 3-fold coordinated sulfur atoms, often present in 
g-GeS$_2$ (13.95 $\%$ of the S particles in the present system). This explains the increase of this feature 
in Ge-rich compositions, as well as in high-pressure samples, as found experimentally.\\
Analyzing the ring size distribution we find that in the process leading from the crystal
to the glass a large proportion of 3-membered rings disappears giving rise to larger rings and an
excess of 2-membered rings. These rings manifest themselves in the vibrational properties of glassy GeS$_2$, 
and principally 2 modes are found, at 200 cm$^{-1}$ and 440 cm$^{-1}$, respectively assigned to bond-bending and 
bond-stretching vibrations of edge-sharing tetrahedra.\\
In the region experimentally assigned to the A$_1$ and A$_{1c}$ modes we find, as 
expected, symmetric-stretch vibrations of corner and edge-sharing 
Ge(S$_{1/2})_4$ tetrahedra respectively, but also vibrations of tetrahedra containing bond defects 
which allows us to explain why these modes do not disappear in Ge-rich systems. \\

{\bf Acknowledgments}\\
We thank David Drabold for providing some of the codes
necessary to analyze the vibrational properties of our sample. Part of the numerical
simulations were done at the ``Centre Informatique National de l'Enseignement
Sup\'erieur'' (CINES) in Montpellier.\\  

\hrule

\begin{figure}
\vspace*{1.5cm}
\centerline{\includegraphics[width=8cm]{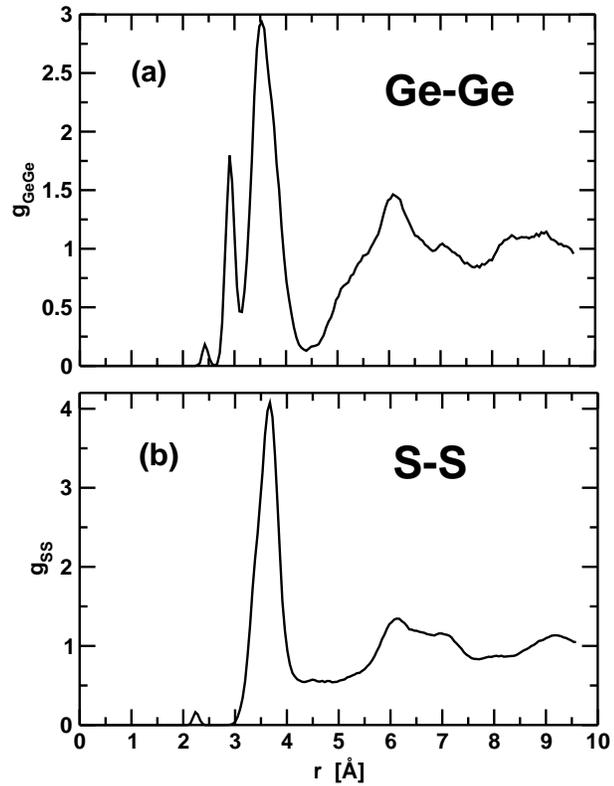}}
\caption{Radial pair correlation functions for Ge-Ge $(a)$ and S-S $(b)$ pairs}
\label{fig1}
\end{figure}

\begin{figure}
\centerline{\includegraphics[width=11cm]{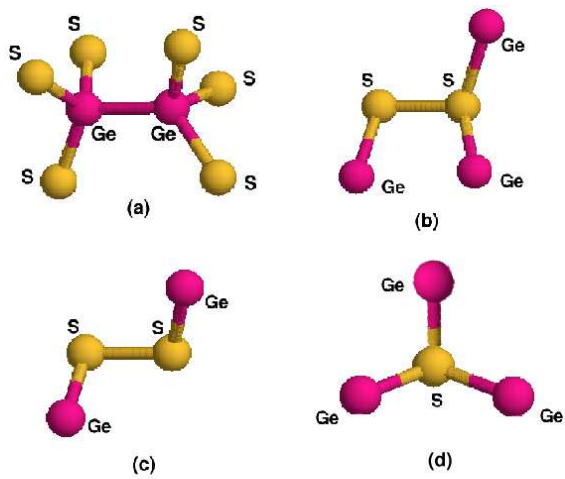}}
\vspace*{-3cm}
\caption{Illustration of three kind of units containing homopolar bonds:
An ethane-like unit presenting a Ge-Ge homopolar bond (a), and two S-S homopolar bonds including one $(b)$ or two $(c)$ non-bridging Sulfur atoms. In addition a 3-coordinated
sulfur atom is illustrated in $(d)$. }
\label{fig2}
\end{figure}
\newpage
\begin{figure}
\vspace*{-0.3cm}
\centerline{\includegraphics[width=15cm]{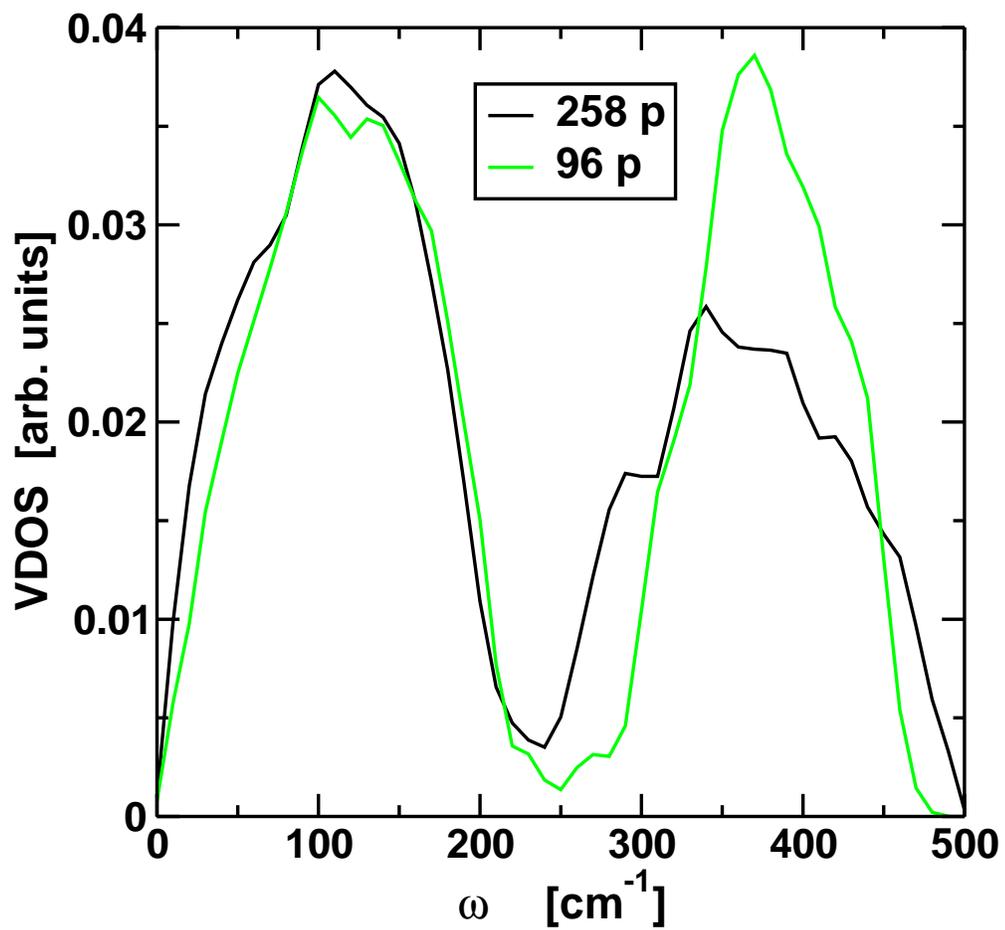}}
\caption{Vibrational density of states (VDOS) of the 258-particles sample compared to the 
one previously obtained for a 96-particle sample~\cite{blaineau}}
\label{fig3}
\end{figure}

\newpage

\begin{figure}
\vspace*{-0cm}
\centerline{\includegraphics[width=8cm]{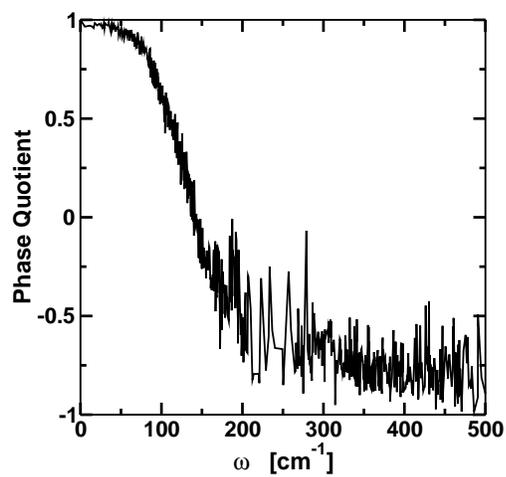}}
\caption{Calculated phase quotient (see text for definition) as a function of $\omega$}   

\label{fig4}
\end{figure}

\begin{figure}
\centerline{\includegraphics[width=8cm]{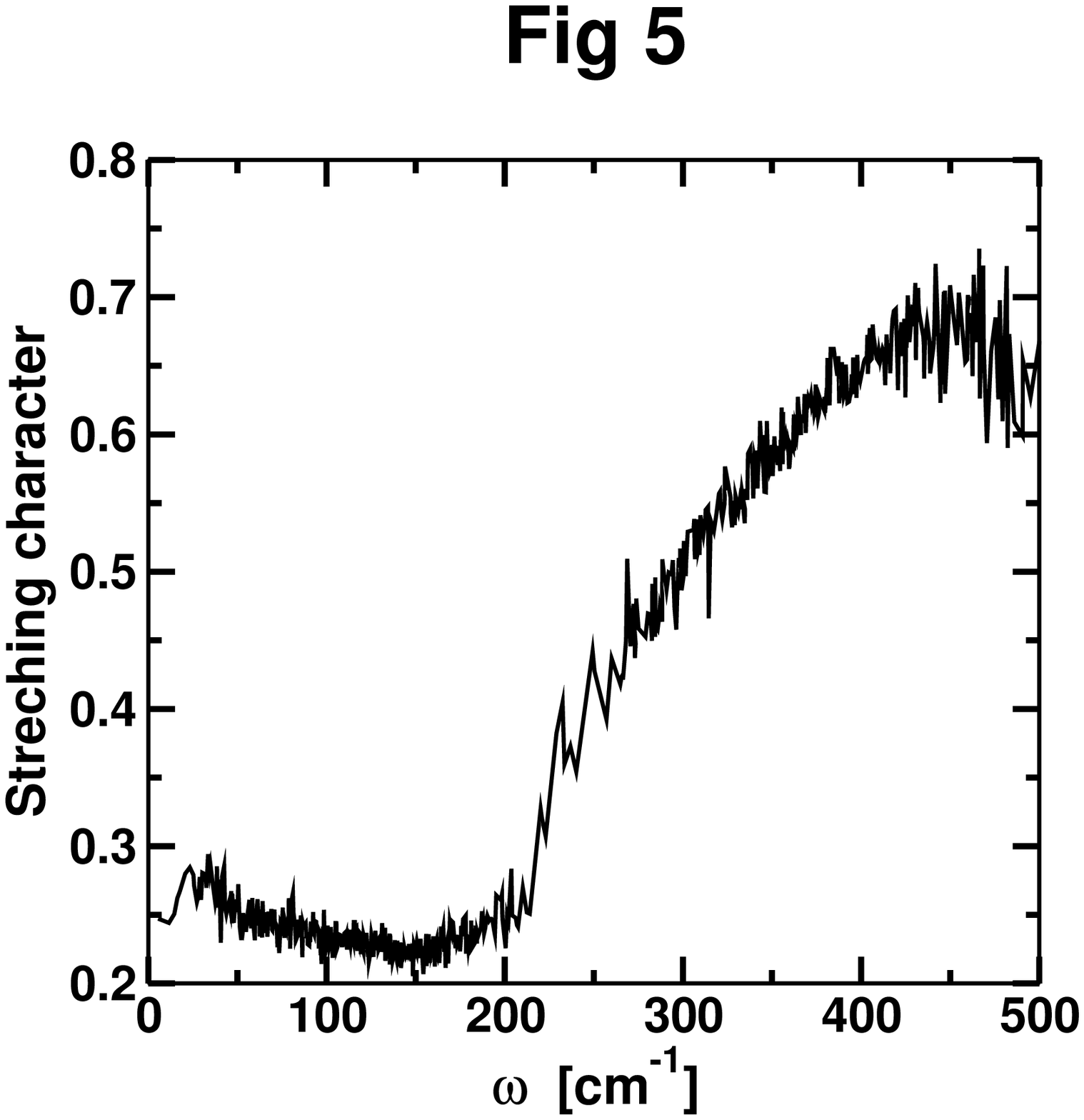}}
\caption{Calculated stretching character (see text for definition) as a function of $\omega$}

\label{fig5}
\end{figure}
\newpage

\begin{figure}
\centerline{\includegraphics[width=15cm]{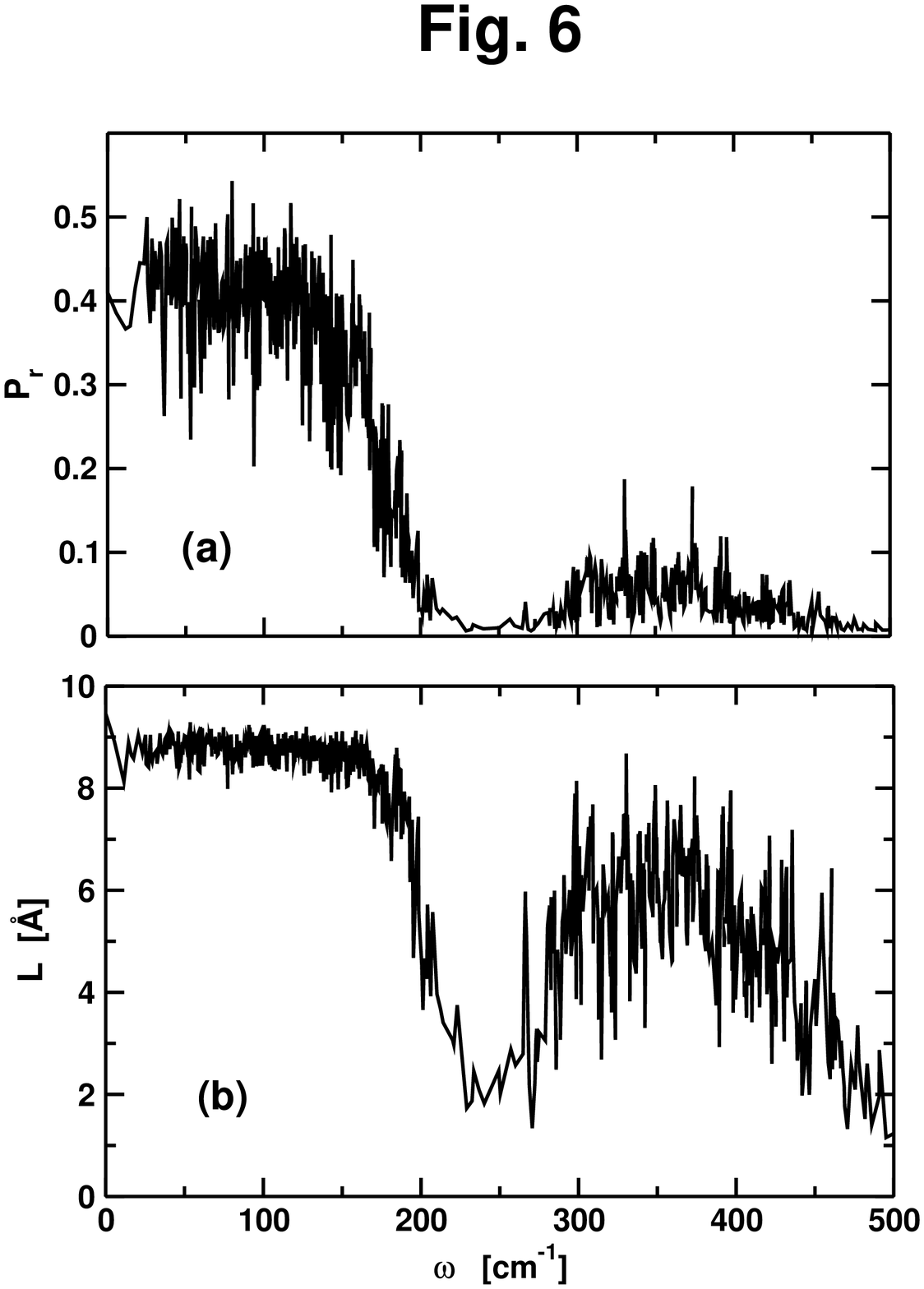}}
\caption{Calculated participation ratio $(a)$ and localization length $(b)$ (see text for definition) 
as a function of $\omega$}
\label{fig6}
\end{figure}
\newpage
\begin{figure}
\centerline{\includegraphics[width=15cm]{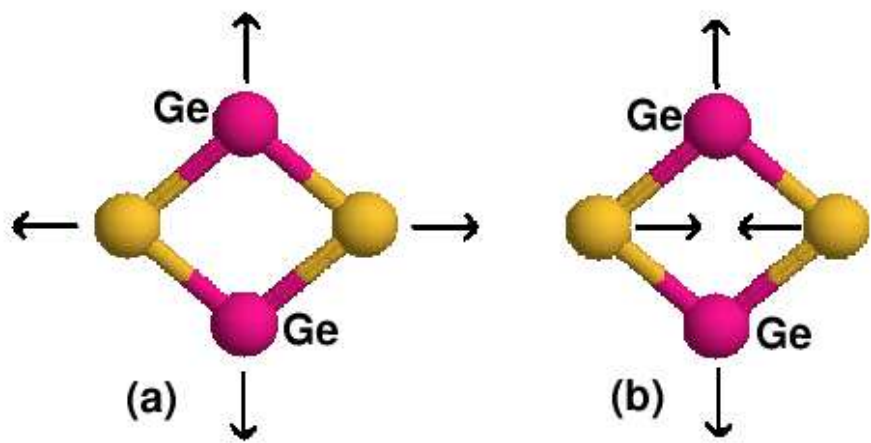}}
\caption{Atomic displacements involved in the vibrational modes at 440 cm$^{-1}$$(a)$ and 200 cm$^{-1}$$(b)$ concerning edge-sharing units}
\label{fig7}
\end{figure}

\begin{figure}
\vspace{-12cm} 
\centerline{\includegraphics[width=18cm]{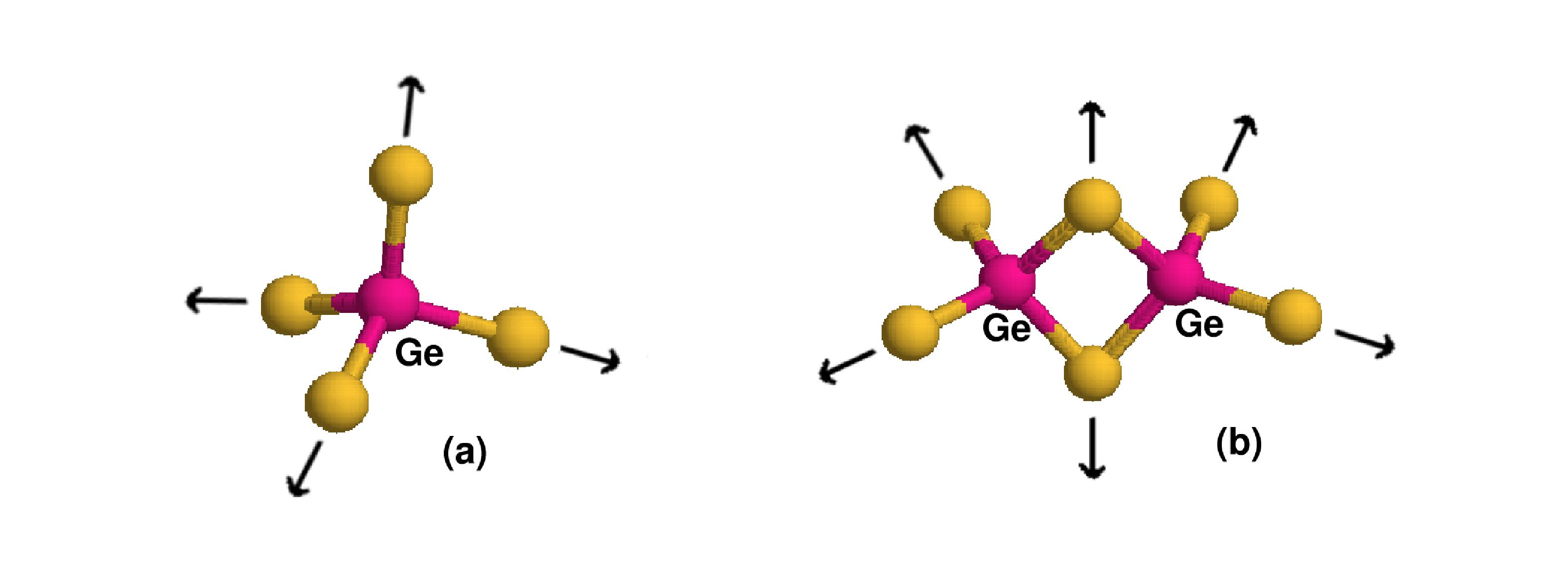}}
\caption{Atomic displacements involved in the A$_1$ $(a)$ and A$_{1_c}$ $(b)$ modes presenting 
symmetric-stretching vibrations of corner-sharing and edge-sharing tetrahedra}
\label{fig8}
\end{figure}


\vskip 0.5cm
\begin{thebibliography}{1}
\bibitem{nemanich} R. J. Nemanich, Phys. Rev. B, {\bf 16}, 1655 (1977).
\bibitem{sugai} S. Sugai, Phys. Rev. B {\bf 35}, 1345 (1987).
\bibitem{bose} M. Yamaguchi, T. Nakayama, and T. Yagi, Phys. Rev. B, {\bf263-264}, 258-260 (1999). 
\bibitem{boolchand} P. Boolchand, J. Grothaus, M. Tenhover, M. A. Hazle, and R. K. Grasselli, Phys. Rev. B {\bf 33}, 5421-5434 (1986).
\bibitem{perakis} A. Perakis, I. P. Kotsalas, E. A. Pavlatou, and C.Raptis, Phys. Stat. Sol. (b) {\bf 211}, 421 (1999).
\bibitem{bychkov} E. Bychkov, M. Fourmentin, M. Miloshova and C. Benmore, ILL Millennium Symposium (Grenoble, France, April 6-7, 2001), p. 54.
\bibitem{transfer} D. Foix, Phd. Thesis, Universit\'e de Pau, France (2003). 
\bibitem{blaineau}  S. Blaineau, P. Jund and D. Drabold, Phys. Rev. B {\bf 67}, 094204 (2003).
\bibitem{fireball} A. A. Demkov, J. Ortega, O. F. Sankey, and M. Grumbach, Phys. Rev B {\bf 52}, 1618 (1995). 
\bibitem{sankey} O. F. Sankey an D. J. Niklewski, Phys. Rev. B, {\bf 40}, 3979 (1989).
\bibitem{dft} P. Hohenberg and W. Kohn, Phys. Rev {\bf 136}, B864 (1964).
\bibitem{lda} D. M. Ceperley and B. J. Alder, Phys. Rev. Lett. {\bf 45}, 566 (1980).
\bibitem{pseudo} G. B. Bachelet, D. R. Hamman, and M. Schluter, Phys. Rev. B {\bf 26}, 4199 (1982).
\bibitem{harris} J. Harris, Phys. Rev. B, {\bf 31}, 1770 (1985).
\bibitem{drabold} M. Durandurdu, D. A. Drabold, and N. Mousseau, Phys. Rev B, {\bf 62} 15307 (2000).
\bibitem{junli} J. Li, D.A. Drabold, Phys Rev. B, {\bf 64} 104206 (2001).
\bibitem{cai}L. Cai and P. Boolchand, Philos. Mag. B {\bf 82}, 1649 (2002).
\bibitem{petri}I. Petri and P. S. Salmon, J. Non-Cryst. Solids {\bf 293-295}, 169 (2001).
\bibitem{expe}  A. Ibanez, M. Bionducci, E. Philippot, L. Desc\^otes, 
R. Bellissent, J. Non-Cryst. Solids {\bf 202}, 248 (1996).
\bibitem{ribes} J. Olivier-Fourcade, J. C. Jumas, M. Ribes, E. Philippot and M. Maurin, Journal of Solid State Chemistry {\bf 23}, 155 (1978).
\bibitem{popovic} Z. V. Popovic, Phys. Lett. {\bf 94A}, 242 (1982).
\bibitem{cluster} I. Fejes, F. Billes, Int. J. Quantum Chem. {\bf 85}, 85 (2001).
\bibitem{bell1} R. J. Bell and D. C. Hibbins-Butler, J. Phys. Chem. {\bf 68}, 2926 (1964).
\bibitem{zotov} M. Marinov and N. Zotov, Phys. Rev. B, {\bf 55}, 2938 (1997).
\bibitem{bell2} R. J. Bell, Methods Comput. Phys. {\bf 15}, 215 (1976).
\bibitem{lucovsky2} G. Lucovsky, F. L. Galeener, R. C. Keezer, R. H. Geils, and H. A. Six, Phys. Rev. B {\bf 10}, 5134 (1974).
\bibitem{lucovsky1} G. Lucovsky, R. J. Nemanich, and F. L. Galeener, in {\em Proceedings of the 7th International Conference on Amorphous and Liquid Semiconductors, Edinburgh, Scotland, 1977}, edited by W. E. Spear(G. G. Stevenson, Dundee, Scotland, 1977), p.130.
\bibitem{jackson} K. Jackson, A. Briley, S. Grossman, D. V. Porezag, and M. R. Pederson, Phys. Rev. B {\bf 60}, R14985 (1999).
 \bibitem{weinstein} B. A. Weinstein and M. L. Slade, in {\em Optical Effects in Amorphous Semiconductors (Snowbird, Utah, 1984)}, edited by P. C. Taylor and S. G. Bishop (AIP, New York, 1984), p.457; see also B. A. Weinstein and R. Zallen, in {\em Light Scattering in Solids IV}, Vol. 54 of {\em Topics in Applied Physics}, edited by M. Cardona and G. Guntherodt (Springer, New York, 1984), p.465.
\bibitem{s8exp} A. Ibanez, M. Bionducci, E. Philippot, L. Desc\^otes, R. Bellissent, Journal of Non-Crystalline Solids {\bf 202}, 248 (1996).



\end{thebibliography}
\end{document}